\documentclass[aps,prc,longbibliography,nofootinbib,twocolumn,floatfix,10pt]{revtex4-1}
\usepackage{blindtext}
\usepackage{sidecap}
\usepackage{subcaption}
\usepackage{physics}
\usepackage{amsmath}
\usepackage{amssymb}
\usepackage[]{graphicx}
\usepackage{hyperref}
\usepackage{float}
\usepackage{blkarray}
\usepackage{bbold}

\usepackage{tikz}
\begin{document}

\title{Imaginary Time Mean-Field Method for Collective Tunneling}
\author{Patrick McGlynn}\email{patrick.mcglynn@anu.edu.au}
\affiliation{Department of Theoretical Physics and Department of Nuclear Physics, Research School of Physics, The Australian National University, Canberra ACT 2601, Australia}
\author{C\'edric Simenel}\email{cedric.simenel@anu.edu.au}
\affiliation{Department of Theoretical Physics and Department of Nuclear Physics, Research School of Physics, The Australian National University, Canberra ACT 2601, Australia}
\date{\today}

\begin{abstract}
\edef\oldrightskip{\the\rightskip}
\begin{description}
\rightskip\oldrightskip\relax
\setlength{\parskip}{0pt} 
\item[Background] 
Quantum tunneling in many-body systems is the subject of many experimental and theoretical studies in fields ranging from cold atoms to nuclear physics. 
However, theoretical description of quantum tunneling with strongly interacting particles, such as nucleons in atomic nuclei, remains a major challenge in quantum physics. 
\item[Purpose] 
An initial-value approach to tunneling accounting for the degrees of freedom of each interacting particle is highly desirable.
\item[Methods] 
{
Inspired by existing methods to describe instantons with periodic solutions in imaginary time, 
we investigate the possibility to use an initial value approach to describe tunneling at the mean-field level.}
Real-time and imaginary-time Hartree dynamics are compared to the exact solution in the case of two particles in a two-well potential. 
\item[Results] 
Whereas real-time evolutions exhibit a spurious self-trapping effect preventing tunneling in strongly interacting systems, the imaginary-time-dependent mean-field method predicts tunneling rates in excellent agreement with the exact solution. 
\item[Conclusions] 
Being an initial-value method, {
it could be more suitable than approaches requiring periodic solutions to describe realistic systems such as heavy-ion fusion. }
\end{description}
\end{abstract}\maketitle

\section{Introduction}
Quantum tunneling, allowing an object to pass a potential energy barrier, even when
classically it does not have enough energy, is one of the most striking concepts of quantum physics. 
In Nature, it produces energy in stars via nuclear fusion, it heats the Earth's interior by $\alpha-$decay,
 it produces new elements in nucleosynthesis, and it is believed to cause DNA
mutations, ageing and cancer. Quantum tunneling also underpins many technological
applications such as the tunneling electron microscope, FLASH memory, tunnel junctions
in solar cells, and tunnel diodes in high speed devices.
Although well-understood for simple systems such as
electrons, the description of quantum tunneling for most systems (e.g., molecules, atomic
nuclei, Bose-Einstein condensates) remains challenging due to their composite
nature and the interaction between their constituents. 

Several methods have been developed in recent years to describe tunneling of interacting particles. 
Due to the complexity of the problem, these methods are either limited to few-body systems \cite{ahsan2010,hunn2013,rontani2013,lundmark2015,dobrzyniecki18} or they require approximations
to the quantum many-body dynamics \cite{umar2009,Hagino2012,simenel2013,fasshauer2016,wen2017,simenel2017,godbey2019b,lode2020}.
As a result, mean-field driven non-exponential decay \cite{zhao2017,alcala2017,potnis2017} as well as
substantial deviation from mean-field dynamics \cite{mclain18} were found in cold atoms systems. 
Pairing effects have also been demonstrated in systems of two and three atoms \cite{zurn2012,zurn2013,gharashi2015,lundmark2015,dobrzyniecki18,dobrzyniecki19}.
Indeed, two atoms can tunnel as a correlated pair when the interaction is strong and attractive \cite{lundmark2015}.
Similar cluster tunneling is found in nuclear systems, e.g., in $\alpha-$decay. 
Despite being composed of four strongly interacting nucleons,  $\alpha-$clusters can be approximated 
as inert particles due to the large difference between their ground-state and first excited state, 
thus reducing theoretical description of $\alpha-$decay to a single-particle tunneling problem. 
However, the situation is much more complicated with heavier systems which can encounter  
non-trivial many-body dynamics while tunneling, as, e.g., in spontaneous fission and in low-energy heavy-ion fusion reactions.
Unlike cold atoms which tunnel sequentially or as small clusters, all nucleons are usually tunneling together as self-bound systems in such reactions.
However, unlike in $\alpha-$decay, individual nucleonic degrees of freedom need to be accounted for. 
Indeed, microscopic nuclear dynamics calculations (see \cite{simenel2018} for a review) show that fusion 
is affected by nucleon transfer \cite{vo-phuoc2016,godbey2017} 
leading to dissipation \cite{Williams2018,simenel2020} and potentially decoherence \cite{dasgupta2007} in many-body tunneling.

At present, there is no model of quantum many-body tunneling for strongly interacting
 systems such as nuclei that explicitly accounts for effects of
 dissipation or decoherence that are induced by nucleonic degrees of
 freedom.
Nevertheless, early works based on instantons and path integral description of quantum mechanics 
opened the possibility for mean-field description of tunneling based on imaginary-time techniques \cite{Levit1980a,Reinhardt1981,pudu1987,Arve1987,Negele1989,Skalski2008}.
Direct implementation of this method, however, is challenging due to the difficulty to find quantum many-particle closed trajectories in imaginary 
time.
To overcome this difficulty, we propose an initial value approach to describe tunneling through 
an imaginary time mean-field 
evolution akin to the way standard real-time mean-field evolution
 techniques such as time-dependent Hartree-Fock (TDHF) can be used to study vibrations (without
 requantization) (see, e.g., \cite{simenel2009,avez2013}). 
This approach is tested in a simple toy-model where 
an exact solution exists, and where the evolution in classically forbidden region can be easily visualised. 
In particular, we show how tunneling probabilities can be extracted at the mean-field level.  
These predictions are compared with the exact solution for various strength of the interaction. 
The toy model has been chosen so that generalisations to more realistic systems are in principle feasible. 

The two-well toy model is described in section~\ref{sec:toy}, where both exact and real-time mean-field dynamics are studied. 
The path integral approach and its application within the toy model are discussed in section~\ref{sec:path}.
The method to compute tunneling probabilities is introduced in section~\ref{sec:proba}.
Potential extensions and applications to more realistic systems are discussed in section~\ref{sec:realistic}, before we conclude in section~\ref{sec:conclusions}.

\section{Two-well model\label{sec:toy}}

We consider a simple toy model with two interacting distinguishable particles evolving according to the Hamiltonian 
\begin{equation}
\hat{H}(1,2)=\hat{h}_0(1)+\hat{h}_0(2)+\hat{v}(1,2).
\end{equation}
The single-particle Hamiltonian is written 
$$\hat{h}_0=\alpha \ket{-}\bra{-},$$ where $\ket{-}$ is the excited state. Its ground-state $\ket{+}$ has   energy 0.
This corresponds to a two-well potential with two possible positions, $\ket{L}$ and $\ket{R}$, for a particle in the left and right well, respectively. These states are related to the  eigenstates of $\hat{h}_0$ by $\ket{\pm}=\frac{1}{\sqrt{2}}(\ket{L}\pm\ket{R})$.
In this model, a particle initially in one well can tunnel to the other well through a potential energy barrier decreasing with $\alpha$. 
The interaction is assumed to occur when both particles are in the same well with $\hat{v}(1,2)=\mu\,(\ket{LL}\bra{LL}+\ket{RR}\bra{RR})$, where $\mu$ is a parameter controlling the interaction strength.
The initial state is chosen as $\ket{LL}$.

It is possible to extend this toy model by considering identical particles (bosons or fermions), adding more particles and modes, and using a more realistic interaction. 
Nevertheless, despite its simplicity this model accounts for the essential aspects of tunneling of interacting particles. 
Moreover, such simplifications allow for easier visualisation of the configuration space available to the system, and thus it gives us a better understanding of its dynamics.
Most importantly, this model is exactly solvable analytically, thus providing a benchmark to test various approximations. 

\subsection{Exact dynamics}

The exact evolution is determined from the time-evolution operator $\hat{U}=e^{-i\hat{H}t}$ (we set $\hbar=1$).
The Hamiltonian of the two-well  model is 
\begin{eqnarray}
	\hat{H}&=&\alpha\left(\ket{-}_1\bra{-}_1\mathbb{1}_2+\mathbb{1}_1\ket{-}_2\bra{-}_2\right)\nonumber\\
	&&+\mu\left(\ket{LL}\bra{LL}+\ket{RR}\bra{RR}\right)\nonumber
\end{eqnarray} 
with $\ket{-}=\frac{1}{\sqrt{2}}(\ket{L}-\ket{R})$. The indices refer to particles 1 and 2. 
In the $\{L,R\}$ basis it is expressed as
\begin{equation}\hat{H}=\,\,\begin{blockarray}{cccc}
	\ket{LL} & \ket{LR} & \ket{RL} & \ket{RR}  \\
	\begin{block}{(cccc)}
		\alpha +\mu& -\alpha/2 & -\alpha/2 & 0  \\
		-\alpha/2 & \alpha & 0 & -\alpha/2  \\
		-\alpha/2 & 0 & \alpha & -\alpha/2  \\
		0 & -\alpha/2 & -\alpha/2 & \alpha+\mu  \\
	\end{block}
\end{blockarray}\,\,\,\,\,.\nonumber\end{equation}
In this basis, the operator counting the particles in the left well is 
\begin{equation}
	\hat{N}_L= \begin{pmatrix}
	2&0&0&0\\
	0&1&0&0\\
	0&0&1&0\\
	0&0&0&0
	\end{pmatrix}.\nonumber
\end{equation}
With the  condition that the particles are initially in $\ket{LL}$, the state of the system at time $t$ is 
$$\ket{\Psi}(t)=\exp(-i\hat{H}t)\ket{LL}.$$ 
The expectation value of $\hat{N}_L$ in this state can be expressed as
\begin{equation}\label{eq:Nlcount}
\langle\hat{N}_L\rangle=1+\frac{\beta-\mu}{2\beta}\cos(\frac{\beta+\mu}{2}t)+\frac{\beta+\mu}{2\beta}\cos(\frac{\beta-\mu}{2}t),
\end{equation}
where $\beta=\sqrt{4\alpha^2+\mu^2}$.
Note that $\langle\hat{N}_L\rangle$ does not depend on the sign of $\mu$, i.e., if the interaction is attractive ($\mu<0$) or repulsive ($\mu>0$). 
 From now on, we set $\alpha=1$, so that the only parameter is the interaction strength $\mu$. 

\begin{figure}[t]
\includegraphics[width=8.5cm]{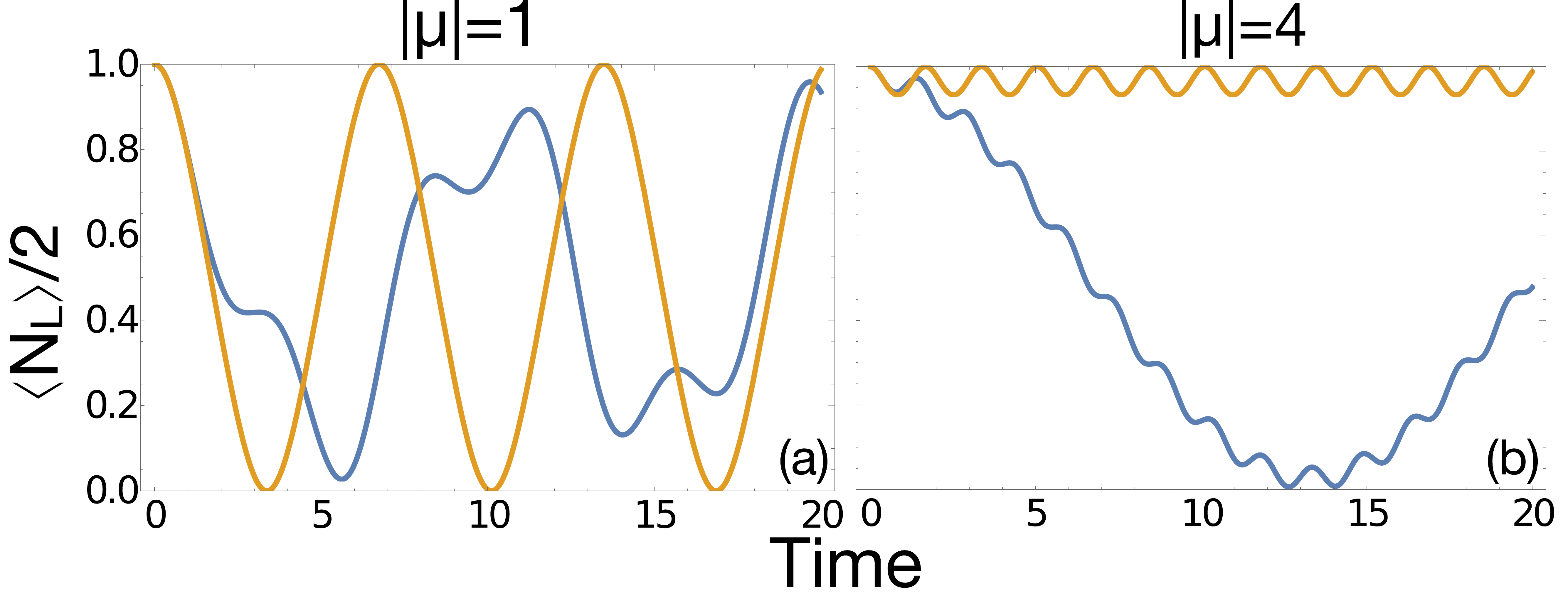}
\caption{
Exact solutions (blue solid lines) and real-time mean field predictions (orange solid lines) of $\langle\hat{N}_L\rangle(t)/2$ for (a) a weak interaction with $|\mu|=1$ and (b) a strong interaction with $|\mu|=4$. Time unit is fixed by our choice of $\alpha=1$.}
\label{fig:2part}
\end{figure}

 The exact evolution $\langle\hat{N}_L\rangle(t)$ is represented in Fig. \ref{fig:2part} (blue solid line) for two values of $\mu$.
 The oscillatory behaviour shows that the particles can tunnel from the left well ($\langle\hat{N}_L\rangle/2=1$) to the right well ($\langle\hat{N}_L\rangle/2=0$). The effect of increasing the interaction strength $\mu$ is to slow down tunneling as the particles are found in the right well at a later time. 

\subsection{Real time mean-field evolution}
Approximations to describe the dynamics of quantum many-body systems are often based on the self-consistent mean-field theory, or time-dependent Hartree theory in the case of  distinguishable particles. The latter assumes that the particles remain independent at all times, i.e., with a state $\ket{\Psi}=\ket{\psi_1}\otimes\ket{\psi_2}$. 
The Hamiltonian is then approximated by $\hat{H}_H(1,2)=\hat{h}_H(1)+\hat{h}_H(2)$ with time-dependent single-particle Hartree Hamiltonian 
$$\hat{h}_H(i)=\hat{h}_0(i)+\bra{\psi_j}\hat{v}(1,2)\ket{\psi_j}$$
 where  $i,j=1,2$ and $j\ne i$.
Note that in this form there is no spurious self-interaction.

As both particles are initially in the same state, they  encounter the same mean-field evolution.
As a result, they remain in identical states $\ket{\psi(t)}=L(t)\ket{L}+R(t)\ket{R}$ obeying the time-dependent Hartree equation
\begin{equation}
i\frac{d}{dt}\ket{\psi(t)}=\hat{h}_H(t) \ket{\psi(t)}.\label{eq:RTDH}
\end{equation}
The time-dependence of $\hat{h}_H(t)$ is due to its self-consistency. 
The equations of motion for $L(t)$ and $R(t)$ then become
\begin{subequations}
\begin{align}
i\frac{d}{dt}L&=\frac{\alpha}{2}(L-R)+\mu L|L|^2 \mbox{ and}\label{eq:dL}\\
i\frac{d}{dt}R&=\frac{\alpha}{2}(R-L)+\mu R|R|^2.\label{eq:dR}
\end{align}
\end{subequations}

Let us define new coordinates $\theta\in[-\pi/2,\pi/2]$ and $\phi\in[-\pi,\pi]$:
\begin{equation}
	\theta=\arcsin(|L|^2-|R|^2) \mbox{ and }\phi=\arg(R/L).
\label{eq:coordinates}
\end{equation}
This choice preserves the normalisation  $|L|^2+|R|^2=1$. 
Both sets of coordinates are equivalent up to a global phase. 
From Eqs.~(\ref{eq:coordinates}) and the normalisation condition we can write 
$$L=\sqrt{\frac{1+\sin\theta}{2}}e^{i\phi_L} \mbox{ and } 
R=\sqrt{\frac{1-\sin\theta}{2}}e^{i\phi_R}$$ 
where we have introduced the phases $\phi_{L,R}$. 
According to Eqs.~(\ref{eq:coordinates}), only their difference $\phi=\phi_R-\phi_L$ is relevant. 

Inserting into Eqs.~(\ref{eq:dL}) and~(\ref{eq:dR}) gives
\begin{subequations}
\begin{align}
i\dot\theta\cos\theta-2(1+\sin\theta))\dot\phi_L=&1+\sin\theta-e^{i\phi}\cos\theta \nonumber\\
&+\mu(1+\sin\theta)^2\label{eq:d1}\\
-i\dot\theta\cos\theta-2(1-\sin\theta))\dot\phi_R=&1-\sin\theta-e^{-i\phi}\cos\theta \nonumber\\
&+\mu(1-\sin\theta)^2,\label{eq:d2}
\end{align}
\end{subequations}
where we have set $\alpha=1$.
Taking the imaginary part gives the first differential equation
\begin{equation}
\dot{\theta}=-\sin\phi\label{eq:thetadot}.
\end{equation}
Taking the real part of Eqs.~(\ref{eq:d1}) and~(\ref{eq:d2}) and rearranging gives the second differential equation
\begin{equation}
	\dot{\phi}=\tan\theta\cos\phi+\mu\sin\theta.\label{eq:phidot}
\end{equation}
Equations (\ref{eq:thetadot}) and (\ref{eq:phidot}) provide a closed set of  equations for the real-time mean-field dynamics of the system.

Solving these equations numerically with initial condition $(\theta,\phi)=(\frac{\pi}{2},0)$ corresponding to both particles in the left well, 
we get the real-time mean-field prediction for $\langle N_L \rangle(t) = 1+\sin\theta(t)$ plotted in Fig.~\ref{fig:2part} (orange solid line).
As in the exact case, the latter does not depend on the sign of $\mu$. 
Apart for the non-interacting case $\mu=0$ (for which mean-field dynamics is obviously exact), we see that mean-field predictions rapidly deviate from the exact solution. 
Although for $|\mu|\leq2$, which we loosely refer to as the ``weakly'' interacting regime, tunneling is observed in the mean-field solution, a transition appears at $|\mu|=2$, above which (``strongly'' interacting regime) the particles are ``trapped" in one well, unable to tunnel completely to the other well. This spurious phenomenon, called ``discrete self-trapping" \cite{Eilbeck1985,Smerzi1997,Milburn1997a}, illustrates the inability of real-time mean-field theory to describe tunneling dynamics in strongly interacting systems.

\begin{figure*}
\centering
\includegraphics[width=18cm]{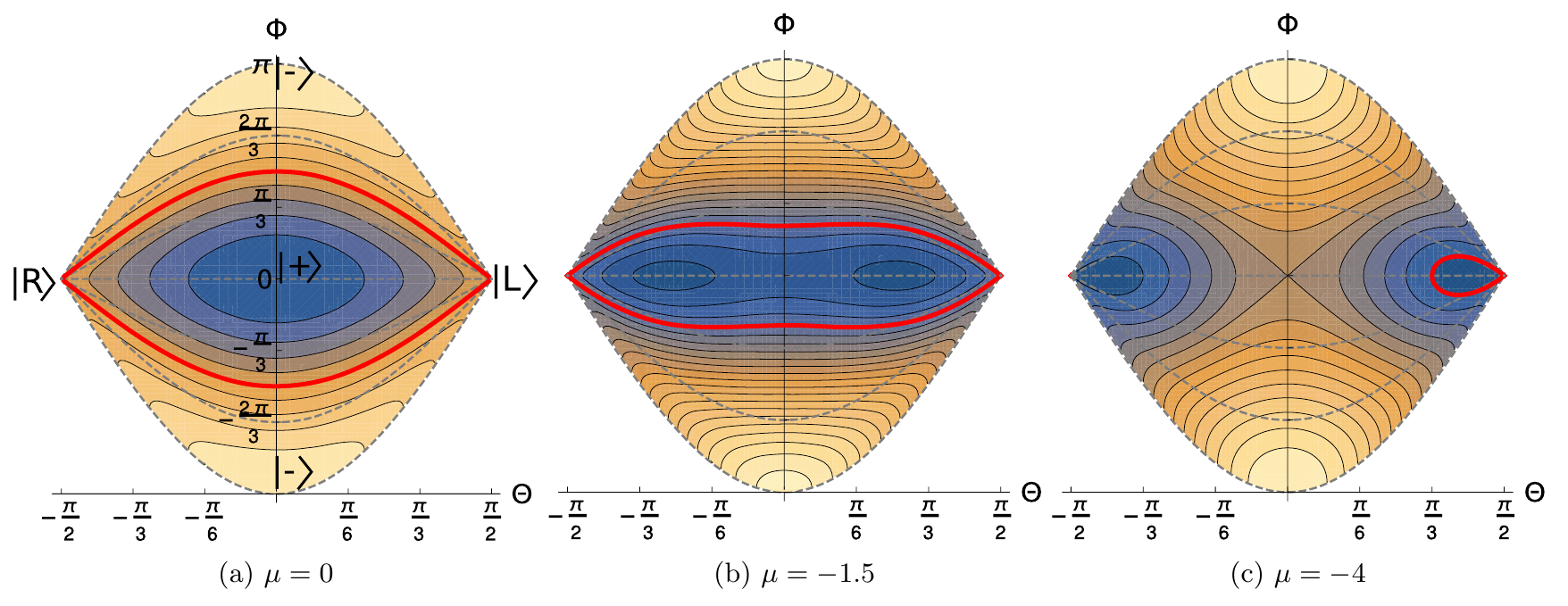}
\caption{
Mean-field energy  contours are shown for (a) free particles, (b) a weak attraction, and (c) a strong attraction. Energy increases from blue to yellow.
Solid red lines show mean-field trajectories starting from the left well. The dashed lines represent constant values of $\phi$.}
\label{fig:enconts}
\end{figure*}

\subsection{Hartree energy}

In order to investigate the origin of self-trapping, let us first determine the energy of the system in the mean-field approximation. 
Without interaction, the total energy is given by 
\begin{eqnarray}
K&=&\sum_{i=1}^{2}\bra{\psi_i}\hat{h}_0\ket{\psi_i}=
\begin{pmatrix}L^*&R^*\end{pmatrix}
\begin{pmatrix}
1&-1\\
-1&1
\end{pmatrix}
\begin{pmatrix}L\\R\end{pmatrix}\nonumber\\
&=&1-2\text{Re}[L^*R]=1-\cos\theta\cos\phi,\nonumber
\end{eqnarray}
where we have used  $\ket{\psi_{1,2}}=L\ket{L}+R\ket{R}$,  Eqs.~(\ref{eq:coordinates}) and 
\begin{equation}\hat{h}_0=\frac{\alpha}{2}
\begin{pmatrix}
		1 & -1   \\
		-1 & 1 \\
\end{pmatrix}\nonumber
\end{equation}
in the $\{\ket{L},\ket{R}\}$ basis, with our choice of $\alpha=1$.

With an interaction treated at the mean-field level, an additional term $U=\frac{1}{2}\sum_{i,j\ne i}\bra{\psi_i\psi_j}\hat{v}(1,2)\ket{\psi_i\psi_j}$ contributes to the total energy $E=K+U$.
Using $\hat{v}(1,2)=\mu\,(\ket{LL}\bra{LL}+\ket{RR}\bra{RR})$, we get 
\begin{eqnarray}
U&=&\mu\begin{pmatrix}L^*&R^*\end{pmatrix}\begin{pmatrix}
|L|^2&0\\
0&|R|^2
\end{pmatrix}\begin{pmatrix}L\\R\end{pmatrix}\nonumber\\
&=&\mu(|L|^4+|R|^4)
=\frac{\mu}{2}\left[1+\sin^2\theta\right].\nonumber
\end{eqnarray}
As a result, the total (Hartree) energy is expressed as
\begin{equation}
	E=1+\frac{\mu}{2}(1+\sin^2\theta)-\cos\theta\cos\phi\label{eq:totalen}
\end{equation}
which is conserved under Eqs. (\ref{eq:thetadot}) and (\ref{eq:phidot}). 

\subsection{Self-trapping}
The self-trapping phenomenon can be understood by examining the mean-field dynamics of the system in the configuration space. 
The latter is entirely defined by the coordinates $\theta$ and $\phi$, allowing for a simple two-dimensional representation as in Figure~\ref{fig:enconts}(a) showing the position of the $\ket{L}$, $\ket{R}$ and $\ket{\pm}$ states. 

Figuress~\ref{fig:enconts}(a-c) show contour plots of the Hartree energy in configuration space
for various interaction strengths. 
As in the exact case, mean-field dynamics conserve total energy. The system is thus bound to follow iso-energy contour lines.
We see that with no (Fig.~\ref{fig:enconts}(a)) or ``weak'' attraction (Fig.~\ref{fig:enconts}(b)), the system is able to go from one well to the other following a classically allowed path (thick solid red lines).
However, in the case of ``strong'' attraction (Fig.~\ref{fig:enconts}(c)), there is no iso-energy contour line connecting both wells. The transition is classically forbidden, preventing tunneling and leading to self-trapping. 


Self-trapping occurs when the states at $\theta=\pm\pi/2$ are not connected by any energy contour line in the $(\theta,\phi)$ plane.
According to Eq.~(\ref{eq:totalen}), the energy at $\theta=\pm\pi/2$ is 
\begin{equation}
E_1=E\big|_{\theta=\pm\pi/2}=1+\mu,\nonumber
\end{equation}
while at $\theta=0$ it is
\begin{equation}
E_2(\phi)=E\big|_{\theta=0}=1+\frac{\mu}{2}-\cos\phi.\nonumber
\end{equation}
A condition for the system to tunnel from one well to another in the realtime mean-field dynamics is that there exist a $\phi$ for which $E_2(\phi)=E_1$ 
(otherwise the system is unable to cross the $\theta=0$ line).  
This is only possible for $\mu/2=-\cos\phi$, leading to the condition $|\mu|\leq 2$. 
Self-trapping then occurs when $|\mu|\geq 2$. 
This condition does not depend on the sign of $\mu$. Thus, despite  the fact that the contour lines are different for attractive or repulsive interactions,  self-trapping occurs in both cases for the same magnitude of the interaction strength. 

\subsection{``Weak'' and ``strong'' interactions}

Some comments are in order regarding the distinction between ``weakly''
and ``strongly'' interacting regimes.
The situation in terms of realistic applications will depend on the system.
For cold atoms, the interaction can be tuned experimentally and the
``transition'' region could then be explored. 

For nuclear systems, in particular in the case of fusion, fission and cluster 
decay, the systems are expected to be well in the ``strong'' interaction 
regime. 
In our toy model, the energy splitting $\Delta E$ 
between the ground and first excited state of the exact Hamiltonian is the quantity that drives the tunneling rate.
 Arve {\it et al.} used 
a two-well potential with parameters adjusted to describe a typical fission 
problem, with energy splittings between the quasi-degenerate eigenstates
of the order of $10^{-13}$~MeV for the ground-state, up to $10^{-2}$~MeV near the 
barrier \cite{Arve1987}. In the two-well model we use, the energy splitting $\Delta E$
is $\sim\alpha^2/|\mu|$ for $|\mu|\gg \alpha$. The interaction $\mu$ is of the order 
of the binding energy per nucleon ($\sim8$~MeV). To get similar splitting as 
Arve et al, we would then set  $\alpha\sim10^{-6}$~MeV for the ground state, 
up to $\sim0.3$~MeV near the barrier. In any case, $\alpha$ remains 
smaller than $2|\mu|$ (recall that the transition appears at $|\mu|=2\alpha$), 
indicating a system clearly in the ``strong'' interaction regime.

\section{Path integral approach\label{sec:path}}

Our goal is now to search for an initial-value mean-field based description of the system which would account for tunneling  in the strongly interacting regime. 
Following Feynman's many-path approach to quantum mechanics~\cite{Feynman1948}, the amplitude of probability for the system to go from the state $\ket{i}$ at time $t_i$ to the state $\ket{f}$ at time $t_f$ is written as a path integral
 \begin{equation}\bra{f}\hat{U}(t_f,t_i)\ket{i}=\int D[\sigma]\exp(iS[\sigma]),
 \nonumber\end{equation}
where $\hat{U}(t_f,t_i)$ is the evolution operator associated with the hamiltonian $\hat{H}$ of the system and $S[\sigma]$ is the action for the path $\sigma(t)$  in configuration space between $t_i$ and~$t_f$.

\subsection{Imaginary time-dependent mean-field equation}

Though elegant, this path integral approach is often too complicated in practice and requires approximations such as the stationary phase approximation (SPA). 
For a single-particle following a classical path $\sigma(t)\equiv x(t)$, 
the SPA leads to the stationary action principle $\delta{S}=0$ of classical mechanics in which quantum tunneling is forbidden. 
Nevertheless, the latter can be recovered approximatively through a Wick rotation changing real time to imaginary time $t\rightarrow -i\tau$ \cite{Zichichi1977}.
Its effect is indeed to change the sign of the potential, thus allowing the system to explore classically forbidden regions. 
This approach is formally equivalent to the WKB semiclassical approximation \cite{Holstein1982,Holstein1982a}.

For a many-particle system, mean-field equations can be recovered from the  stationary action principle with the Dirac action 
\begin{equation}
S=\int_{t_i}^{t_f}dt \bra{\Psi}i\partial_t-\hat{H}\ket{\Psi}
\label{eq:Dirac}
\end{equation} 
while restricting the variational space to independent particle states.  
However, as illustrated by our toy model, this theory does not account for tunneling in the strong interaction regime. 
Nevertheless, applying a Wick rotation should produce imaginary-time mean-field equations accounting for tunneling.
Replacing $t\rightarrow-i\tau$ in Eq.~(\ref{eq:RTDH}) 
leads to an imaginary-time dependent Hartree equation 
\begin{equation}
\frac{d}{d\tau}\ket{\psi(\tau)}=-\hat{h}_H(\tau) \ket{\psi(\tau)}.
\end{equation}
As in real time evolution, computing observables $\langle\hat{Q}\rangle(\tau)$ also requires a conjugate state which we now define. 

\subsection{Wave-function of conjugate state}

In real time, the wave-function of a single-particle conjugate state is given by $\tilde{\psi}(x,t)=\bra{\psi(t)}\ket{x}=\psi^*(x,t)$.
It is convenient to write the wave-function as 
\begin{equation}
\psi(x,t)=\sqrt{\rho(x,t)}e^{i\phi(x,t)}, \label{eq:psit}
\end{equation}
with 
\begin{subequations}
\begin{align}
\rho(x,t)&=\tilde{\psi}(x,t)\psi(x,t) \,\,\mbox{ and}\label{eq:rhot}\\
\phi(x,t)&=\frac{1}{2i}\ln\left(\frac{\psi(x,t)}{\tilde{\psi}(x,t)}\right), \label{eq:phit}
\end{align}
\end{subequations}
leading to $\tilde{\psi}(x,t)=\sqrt{\rho(x,t)}e^{-i\phi(x,t)}$.

Imaginary-time evolution is obtained from a Wick rotation $t\rightarrow-i\tau$.
This has a consequence for how the conjugate of a single-particle wave-function is defined. 
The conjugate is used to compute expectation values 
\begin{equation}
\langle\hat{Q}\rangle(\tau)=\int dx \,\,\,\tilde{\psi}(x,\tau) \,Q(x,y)\, \psi(x,\tau)\nonumber
\end{equation}
which are transformed under the Wick rotation as\footnote{\label{foot1}This expression is only correct for a time-independent Hamiltonian for which the imaginary-time evolution operator is given by $\exp(-\hat{H}\tau)$. For a time-dependent Hamiltonian, such as in the self-consistent mean-field approximation, it should be replaced by $T\exp[-\int_0^\tau\hat{H}(\tau)d\tau]$, where $T$ denotes time ordering.}
$$\langle\Psi(0)|e^{i\hat{H}t}\hat{Q}e^{-i\hat{H}t}|\Psi(0)\rangle\longrightarrow \langle\Psi(0)|e^{\hat{H}\tau}\hat{Q}e^{-\hat{H}\tau}|\Psi(0)\rangle.$$
The expectation value is then given by
\begin{equation}
\langle\hat{Q}\rangle(\tau)=\langle\Psi(-\tau)|\hat{Q}|\Psi(\tau)\rangle
\end{equation}
implying
\begin{equation}\tilde{\psi}(x,\tau)=\psi^*(x,-\tau).\label{eq:psitau}\end{equation} 

In imaginary time, $\rho$ and $\phi$ become complex.
It is easy to show from Eqs.~(\ref{eq:psit}),~(\ref{eq:rhot}) and~(\ref{eq:phit}) that $\rho(\tau)=\rho^*(-\tau)$ and $\phi(\tau)=\phi^*(-\tau)$.
As a result, Eq.~(\ref{eq:psitau}) becomes
\begin{equation}
\tilde{\psi}(x,\tau)=\sqrt{\rho(x,\tau)}\exp(-i\phi(x,\tau)).
\end{equation}
As a result, the conjugate in imaginary time has the same structure as in real time.

\subsection{Classically forbidden region and final condition}
Computing expectation values of observables in imaginary time thus requires both forward and backward evolutions. 
These are nevertheless initial value equations as only $|\Psi(0)\rangle$ is required to compute both evolutions in the classically forbidden region. 

Criteria must  be defined for where to ``stop'' the calculation.
One (or several)  observable $O_f$ can be used to define  such criteria. 
The system needs then to be evolved in imaginary time until the condition (see footnote \ref{foot1})
$$O(\tau) = \langle\psi_0| e^{\hat{H}\tau} \hat{O} e^{-\hat{H}\tau} |\psi_0\rangle = O_f$$
 is reached. 
This defines the mean-field path from which the action and then the
 probability (see Sec.~\ref{sec:proba}) to reach $\langle\hat{O}\rangle=O_f$ can be computed. 
 For instance, $\hat{O}$ could be
 the quadrupole operator in fusion/fission problems, or the center of
 mass in cluster decay (see section~\ref{sec:applications}). 
 
 Note that the final state is not necessarily in
 the classically allowed region (in that case, however, a connection to
 real time dynamics cannot be performed).
This is a major difference with earlier implementation of the imaginary-time mean-field approximation~\cite{Levit1980a}
which required bounce solutions with the condition 
$$\ket{\Psi(\tau)}=\ket{\Psi(-\tau)}.$$

\subsection{Application to two-well model}

\begin{figure}[h!]
\includegraphics[width=7cm]{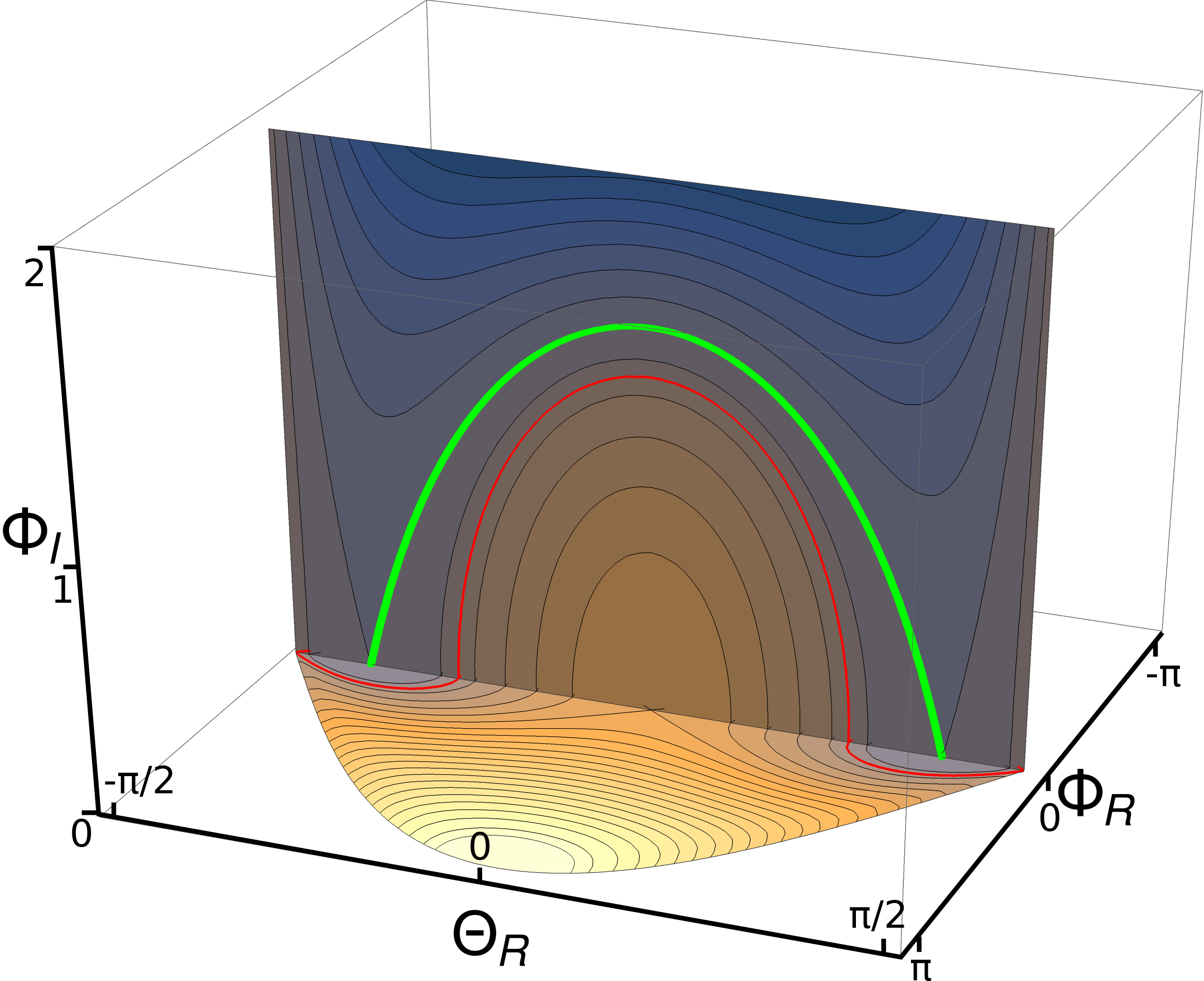}
\caption{The real and imaginary-time mean-field energy is computed with $\mu=-3$ and is increasing from blue to yellow. The $(\theta_R,\phi_R)$ horizontal plane shows the energy for $\phi_I=\theta_I=0$. 
 The $(\theta_R,\phi_I)$ vertical plane shows the energy for $\phi_R=\theta_I=0$.
The red (respectively green) solid line shows an iso-energy contour connecting the left and right wells (resp. the left and right mean-field ground-states).}\label{fig:3contours}
\end{figure}

The equations of motion in imaginary time  are obtained from a Witck rotation of Eqs.~(\ref{eq:thetadot}) and~(\ref{eq:phidot}):
\begin{equation}
\frac{d\theta}{d\tau}=i\sin\phi\,\,\,\mbox{ and }\,\,\,\frac{d\phi}{d\tau}=-i\tan\theta\cos\phi-i\mu\sin\theta.\label{eq:ITEOM}
\end{equation}
The coordinates $\theta=\theta_R+i\theta_I$ and $\phi=\phi_R+i\phi_I$ are now complex. 
The total energy of the system in the imaginary-time-dependent Hartree theory is still conserved and given by Eq.~(\ref{eq:totalen}). 
As a result, as long as the initial condition is in the classically allowed region, i.e., with $\theta_I(t_i)=\phi_I(t_i)=0$, this energy remains real.
This condition, together with the constant norm  $\langle\psi(-\tau)|\psi(\tau)\rangle=1$ also impose relationships between $\theta$ and $\phi$. 

Figure~\ref{fig:3contours} shows this energy for a strongly attractive system. 
The horizontal plane gives the energy for real time evolution, as in Fig.~\ref{fig:enconts}.
The vertical plane represents the energy for imaginary time evolution with $\phi_R=0$.
It is now possible for the system to go from the left well to the right one following a combination of real and imaginary time evolutions (solid red line). 
This demonstrates the ability of imaginary time mean-field evolution to explore classically forbidden region through quantum tunneling.

\section{Tunneling probability\label{sec:proba}}

\begin{figure}[htbp]
	\includegraphics[width=\columnwidth]{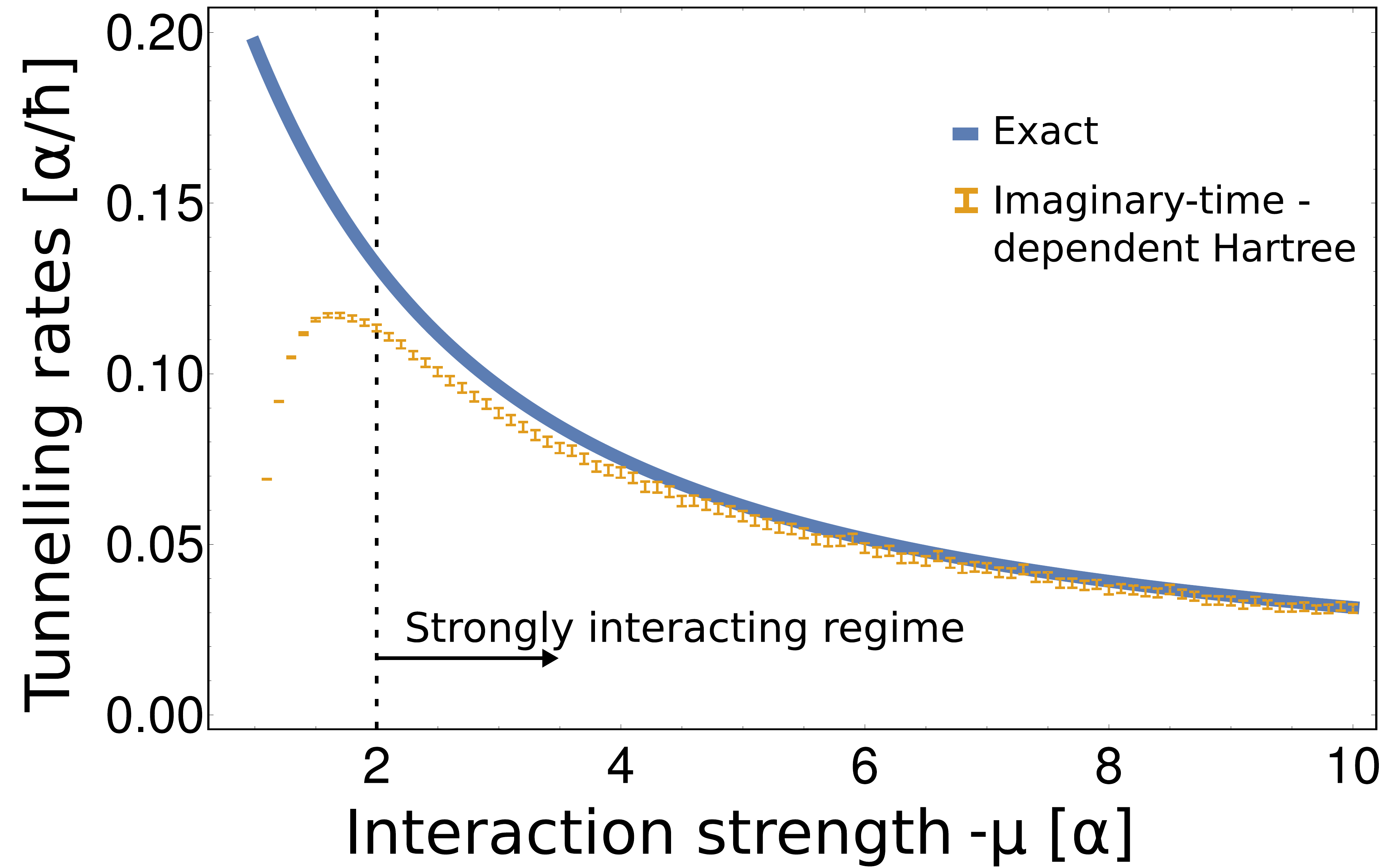}
	\caption{
	Exact (solid line) and imaginary-time-dependent mean-field (symbols) tunneling probabilities per unit of time (in units of $\alpha/\hbar$) are plotted as a function of the attraction strength $-\mu$ (in units of $\alpha$).\label{fig:exactandapprox} Error bars correspond to numerical uncertainty of 1\% in the action. 
	}
\end{figure}

Now that we found mean-field tunneling paths, our next task is to calculate their associated tunneling probability per unit of time 
(tunneling rate) and compare it with the exact case. 

Consider a mean-field evolution from $\ket{i}$ to $\ket{f}$ over a time $T$. 
In real-time, the probability to end up in $\ket{f}$ is  $|\bra{f}\hat{U}(T)\ket{i}|^2=|e^{iS}|^2=1$ as the Dirac action 
in Eq.~(\ref{eq:Dirac}) for this path is real. 
The energy being constant,  the global phase $ET$ is irrelevant. 

In imaginary-time, this probability is now given by $|e^{iW}|^2$ with
\begin{eqnarray}
W(T)&=&-\int_0^Td\tau \bra{\Psi}\partial_\tau\ket{\Psi}\nonumber\\
&=&-\sum_{i=1}^N\int_{0}^T d\tau \bra{\psi_i(\tau)}\pdv{\tau}\ket{\psi_i(\tau)}\nonumber
\end{eqnarray}
where $N$ is the number of particles.
Using results from the previous section, we find
\begin{subequations}
	\begin{align}
	\bra{\psi}\pdv{\tau}\ket{\psi}
	&=\int dx \,\left(\frac{1}{2}\pdv{\rho(x,\tau)}{\tau}+i\rho(x,\tau)\pdv{\phi(x,\tau)}{\tau}\right)\nonumber\\
	&=\int dx\,  i\rho(x,\tau)\pdv{\phi(x,\tau)}{\tau},\nonumber
	\end{align}
\end{subequations}
where we used the fact that $\int dx\, \rho(x,\tau)$ is  constant. 

In the toy model, the space integral $\int dx$ is simply replaced by a discrete sum $\sum_{L,R}$ over the left and right states, giving
\begin{equation}
W[T]=-i\int_0^Td\tau\,\,\left(\rho_L\pdv{\phi_L}{\tau}+\rho_R\pdv{\phi_R}{\tau}\right)\nonumber
\end{equation} 
where 
$$\rho_{L}=\frac{1+\sin\theta}{2} \mbox{  and } \rho_{R}=\frac{1-\sin\theta}{2}.$$
Once again, only the relative phase $\phi=\phi_R-\phi_L$ matters, thus we set $\phi_{R}=\phi/2$ and $\phi_{L}=-\phi/2$, giving
\begin{equation}
W[T]=\frac{i}{2}\int_0^Td\tau\,\,\sin\theta\partial_\tau\phi.\nonumber
\end{equation}
Using the second equation of motion~(\ref{eq:ITEOM}), we finally get
\begin{equation}
W[T]=\frac{1}{2}\int_0^Td\tau\,\,(\tan\theta\cos\phi+\mu\sin\theta)\sin\theta. \nonumber
\end{equation}
Up to an irrelevant global phase,
the tunneling probability amplitude for an imaginary-time-dependent mean-field path $\Psi(\tau)$ connecting the two wells is then given by 
\begin{equation}
e^{iS[\Psi]}\equiv e^{-i\int d\tau\bra{\Psi}\partial_\tau\ket{\Psi}}= e^{\frac{i}{2}\int_0^Td\tau\,(\tan\theta\cos\phi+\mu\sin\theta)\sin\theta}.\nonumber
\end{equation}
As $\theta$ and $\phi$ are now complex quantities, the tunneling probability $|e^{iS[\Psi]}|^2$ associated with this path can be less than one. 

By analogy with the standard semi-classical treatment of $\alpha-$decay \cite{Gamow1928}, the tunneling rate is given by the tunneling probability 
multiplied by the frequency at which the system ``hits'' the potential barrier,
 i.e., the frequency of the oscillation observed in mean-field trapping [see orange line in Fig.~\ref{fig:2part}(b)].
As the system may tunnel from different configurations along a real-time iso-energy contour [see solid red line in Fig.~\ref{fig:enconts}(c)],
the tunneling rate is in principle obtained by averaging over the associated imaginary-time paths.
To a good approximation, this corresponds to the tunneling rate for the path connecting the left and right degenerate mean-field ground-states, indicated by the solid green line in Fig.~\ref{fig:3contours}.

In the exact case, the tunneling rate is simply given by twice the frequency at which the system oscillates between left and right wells.
As shown in Eq.~(\ref{eq:Nlcount}), this oscillation has two modes at $\omega_\pm=\frac{\sqrt{4\alpha^2+\mu^2}\pm|\mu|}{2\pi}$.
Only the lowest frequency is associated with tunneling, giving an exact tunneling rate $2\omega_-=\frac{2}{\pi}\Delta E$ 
where $\Delta E=\frac{\sqrt{4\alpha^2+\mu^2}-|\mu|}{2}\xrightarrow[|\mu|\gg\alpha]{}\frac{\alpha^2}{|\mu|}$ is the energy difference between the ground and first excited states. 

The mean-field and exact tunneling rates are compared in Fig.~\ref{fig:exactandapprox}.
Although the imaginary-time mean-field predictions are wrong for weakly interacting ($|\mu|<2$) systems -- in which case the real-time mean-field prediction can be used anyway -- it is in excellent agreement with the exact case in the strongly interacting regime,
reproducing well the slowing down of tunneling with increased interactions.

\section{Towards  realistic applications\label{sec:realistic}}

The purpose of the two-well model is to illustrate the imaginary time mean-field 
method and compare with an exact solution (which would be hard to obtain 
with more realistic models). This is of course a first step and for the method 
to be useful, its applicability to more realistic systems needs to be demonstrated.

\subsubsection{Cartesian grids}

The toy model has only two states per particle, while the single-particle Hilbert 
space for a one-dimensional discretised cartesian grid has as many states 
as the number of points in the grid --  typically $\sim100$. 
Naturally, numerical simulations with non-unitary evolution operators 
such as $\exp(\pm\hat{H}\tau)$ 
present additional technical challenges in terms of stability and convergence. 
The generalisation from one to three dimensions will be another computational 
challenge, though it does not bring additional formal difficulty in terms of the 
algorithm itself. 

\subsubsection{Spin and exchange terms}

Other extensions include the inclusion of spin and exchange terms. 
Spin $1/2$ degrees of freedom can be accounted for in the same way as in
real-time calculations where each single-particle is treated as a 2-spinor 
$(\varphi_\uparrow(\mathbf{r}), \varphi_\downarrow(\mathbf{r}))$. 
An exchange (Fock) term also appears in the case of identical 
particles. In general, it is non-local and often requires a major extra 
computation cost. However, in the case of contact interactions (often used in cold atoms
systems as well as in nuclear physics, e.g., with the Skyrme effective interaction \cite{skyrme1956}),
the exchange terms are easily accounted for. 
For Coulomb interactions, the exchange term can also be included via the Slater approximation.

\subsubsection{Potential applications\label{sec:applications}}

Several applications could be considered:
\begin{itemize}
\item {\it Interacting particles in an external potential}\\
Cluster dynamics such as $\alpha-$decay or emission of atom clusters
can be studied with an external potential. 
In this case, the external potential simulates the mean-field of the  
particles which do not belong to the cluster, while each particle of the cluster is 
treated explicitly. 
One could study the effect of the internal degrees of freedom of the cluster
while it tunnels as a whole. 
In this case the classically forbidden region could be defined as the 
turning points of the external potential in the usual way. 
\item {\it Merging of two self-bound systems}\\
A typical example is the fusion of two atomic nuclei. Note that the nuclei
are self-bound and thus there is no external potential in this problem, i.e.,
$h_0$ only contains the kinetic energy of the nucleons. The nuclei are 
bound thanks to their strong nuclear interaction. The Coulomb barrier
preventing fusion in the classical case is produced by the competition of
long-range Coulomb repulsion between protons and the short range 
nuclear attraction of all nucleons, both terms being part of the interaction $v(1,2)$. In this
case, the real-time mean-field dynamics is only able to reach fusion if 
the kinetic energy at large distance exceeds the Coulomb barrier height 
(for a central collision). At lower energy, tunneling will be obtained 
through the imaginary time mean-field method. 

As illustrated by the red 
line in Fig.~\ref{fig:3contours}, the system may explore different configurations through 
real-time dynamics, with each of these configurations potentially serving 
as initial condition for the imaginary time evolution. In principle, a weighting
of each possibility should be determined. In practice, however, the 
transmission through the barrier is expected to be dominated by the 
trajectory starting from the distance of closest approach. 
\item {\it Scission of a self-bound metastable system}\\
Self-bound systems can be in a local minimum of their potential energy
surface, with more stable configurations corresponding to disconnected 
fragments. This is the case of fission in heavy nuclei. Here, again, the 
parent nucleus is self-bound and no external potential is required ($h_0$
only contains a kinetic energy term). In the case of spontaneous fission
in particular, all directions in the multidimensional potential energy surface
are classically forbidden, thus the initial condition for the imaginary time
evolution is  well defined\footnote{\label{foot2}In practice, a small deviation from the 
mean-field ground state would be needed to initiate the evolution in the 
right ``direction'' (e.g., a small increase of the quadrupole moment). }. 
\end{itemize}


\section{Conclusions\label{sec:conclusions}}
Theoretical description  of tunneling in strongly interacting systems such as atomic nuclei remains a challenging problem. 
Standard real-time mean-field approaches are unable to account for many-body tunneling due to spurious ``self-trapping''.
Using a simple model with two particles in a two-well potential, we demonstrated the possibility to overcome this limitation by allowing imaginary-time mean-field evolution. 
Tunneling probabilities are in excellent agreement with the exact solution in the strongly interacting regime. 

These results are promising and encourage applications to more realistic systems. 
A first natural extension is to increase the number of modes, e.g., using cartesian grids with one or more dimensions. 
Computational effort only increases linearly with the number of particles at the mean-field level, so simulating tunneling dynamics of larger systems (out of reach to exact and few-body techniques) should not be an issue. 
The imaginary-time mean-field equations could also be extended to include pairing correlations \cite{Levit2020}.\\
 
 \begin{acknowledgments}
We are grateful to R. Bernard for useful discussions.
This work has been supported by the Australian Research Council Discovery Project (project number DP190100256) funding scheme.
\end{acknowledgments}

\twocolumngrid
\bibliography{library}

\end{document}